# Cited References and Medical Subject Headings (MeSH) as Two Different Knowledge Representations: Clustering and Mappings at the Paper Level

Loet Leydesdorff,*[a] Jordan A. Comins,[b] Aaron A. Sorensen,[c]

Lutz Bornmann,[d] and Iina Hellsten[e]

## Abstract

For the biomedical sciences, the Medical Subject Headings (MeSH) make available a rich feature which cannot currently be merged properly with widely used citing/cited data. Here, we provide methods and routines that make MeSH terms amenable to broader usage in the study of science indicators: using Web-of-Science (WoS) data, one can generate the matrix of citing versus cited documents; using PubMed/MEDLINE data, a matrix of the citing documents versus MeSH terms can be generated analogously. The two matrices can also be reorganized into a 2-mode matrix of MeSH terms versus cited references. Using the abbreviated journal names in the references, one can, for example, address the question whether MeSH terms can be used as an alternative to WoS Subject Categories for the purpose of normalizing citation data. We explore the applicability of the routines in the case of a research program about the amyloid cascade hypothesis in Alzheimer's disease (AD). One conclusion is that referenced journals provide archival structures, whereas MeSH terms indicate mainly variation (including novelty) at the research front. Furthermore, we explore the option of using the citing/cited matrix for main-path analysis as a by-product of the software.

---

[a] University of Amsterdam, Amsterdam School of Communication Research (ASCoR), PO Box 15793, 1001 NG Amsterdam, The Netherlands; loet@leydesdorff.net ; * corresponding author.
[b] Center for Applied Information Science, Virginia Tech Applied Research Corporation, Arlington, VA, United States; jcomins@gmail.com
[c] Digital Science, Inc., 1 Canal Park, Suite 1A, Cambridge, MA, 02141 USA; aaron@uberresearch.com
[d] Division for Science and Innovation Studies, Administrative Headquarters of the Max Planck Society, Hofgartenstr. 8, 80539 Munich, Germany; bornmann@gv.mpg.de
[e] Vrije Universiteit Amsterdam, Dep. of Organization Sciences, de Boelelaan 1081, 1081HV Amsterdam, The Netherlands; hellsten.iina@gmail.com

## 1. **Introduction**

The ability to define research fields is one of several great challenges in information science (Chen, 2016). Early efforts relied on classifying publication sources, such as journals, to define research fields. In addition to disciplinary journals, however, the literature databases Web of Science (WoS, Thomson Reuters) and Scopus (Elsevier) contain multi-disciplinary journals such as *Science* and *Nature*. In recent years, new journals which are not organized along disciplinary lines, have been added to the databases. *PLoS ONE*, for example, tends to disturb the existing classifications of journals (Leydesdorff & De Nooy, in press). In response to these changes, bibliometricians have begun to cluster the database at the level of documents instead of journals (e.g., Waltman & van Eck, 2012; cf. Hutchins, Yuan, Anderson, & Santangelo, 2016).

An alternative to clustering documents on the basis of direct citations could be to use databases that are more specialized than WoS and Scopus, but with professional indexing at the document level. The National Library of Medicine, for example, makes a huge investment to maintain a classification system of Medical Subject Headings (MeSH) as tags to the PubMed/MEDLINE database (which is publicly available at http://www.ncbi.nlm.nih.gov/pubmed/advanced).[1] The classification at the article level is elaborated in great detail (Agarwal & Searls, 2009), with a hierarchical tree covering sixteen separate branches that can reach up to twelve levels of depth. Diseases, for example, are classified under C.

---

[1] The National Library of Medicine of the United States (NLM) has constantly received substantial funding for to maintain and update its biomedical and health information services—for example, the 2015 budget for these services was $117 Million (National Library of Medicine, 2015). This has enabled a relatively uniform application of the MeSH classification to publications by indexers over many years (Hicks & Wang, 2011, at p. 292; Petersen *et al*., 2016).

“Alzheimer’s disease” (AD) for example is classified as C10.228.140.380.100 under “Dementia,” as C10.574.945.249 under “Neurodegenerative diseases,” and as F03.615.400.100 under “Neurocognitive disorders” in the F-branch covering “Psychiatry and psychology.” Unlike other disciplinarily specialized databases such as Chemical Abstracts (Bornmann *et al*., 2009), the multiple tree-structure of the *Index Medicus* allows for mapping documents differently across heterogeneous domains (Leydesdorff, Rotolo, & Rafols, 2012; Rotolo, Rafols, Hopkins, & Leydesdorff, 2016). Unlike WoS or Scopus, Medline does not cover the full range of disciplines; but a large part of the scholarly literature in the life sciences is included even more exhaustively than in the more comprehensive databases (Lundberg *et al*., 2006).

A version of MEDLINE is integrated in the databases of Thomson Reuters. The advantage of this installation is that the “times cited” of each record (if the document is also available in the WoS Core Collection of the Citation Indices) is available on screen; but this field is not integrated when the records are downloaded. Rotolo & Leydesdorff (2015) provide software for integrating the “times cited” from the citation indices at WoS into the MEDLINE data. One technical advantage of the installation at PubMed is that the retrieval is not restrained. Using WoS, one can download only 500 records at a time and Scopus has a maximum of 2,000 records.

The MeSH terms attributed to a paper can be considered as references to a body of knowledge stored as documents in a database. Whereas the cited references are provided by the authors themselves, the MeSH categories are attributed by professional indexers. Using MeSH terms as references, one can envisage a matrix of documents referencing MeSH comparable to the cited/citing matrix at the article level. Both cited references and MeSH terms can be considered as attributes of articles, and thus be combined and compared using various forms of multi-variate analysis. The two matrices can also be integrated into a 2-mode matrix of MeSH terms versus cited references. In this brief communication, we explore these options computationally and describe software that has been developed and made available for this purpose on the internet. We discuss the opportunities and the pros and cons of various approaches.

## 2. Methods

### *2.1. Data*

At the professional suggestion of one of us (AS, the scientometrics editor of the *Journal of Alzheimer's Disease*), we selected the amyloid cascade hypothesis in Alzheimer's disease (AD) as a test case to develop software and routines to merge and analyze citation information from the Web of Science and MeSH. The amyloid cascade hypothesis in AD was formulated by Hardy and Allsop in 1991 (cf. Hardy and Higgins, 1992; Selkoe, 1991). Reitz (2012: 1) summarized this hypothesis as follows:

> "Since 1992, the amyloid cascade hypothesis has played a prominent role in explaining the etiology and pathogenesis of Alzheimer's disease (AD). It proposes that the deposition of $\beta$-amyloid (A$\beta$) is the initial pathological event in AD leading to the formation of senile plaques (SPs) and then to neurofibrillary tangles (NFTs), neuronal cell death, and ultimately dementia. While there is substantial evidence supporting the hypothesis, there are also limitations: (1) SP and NFT may develop independently, and (2) SPs and NFTs may be the products rather than the

> causes of neurodegeneration in AD. In addition, randomized clinical trials that tested drugs or antibodies targeting components of the amyloid pathway have been inconclusive."

For the purpose of this study, the search string '("Alzheimer disease"[MeSH Terms] AND "amyloid beta-protein precursor"[MeSH Terms]) AND "mice, transgenic"[MeSH Terms])' was proposed to encompass the relevant literature. This string provided us (on March 6, 2016) with a retrieval of 3,558 records in both PubMed/MEDLINE and the MEDLINE version in WoS. Using PubMed Identifiers (PMID numbers), 3,416 of these records could be retrieved in the WoS Core Collection. As noted, not all journals covered by PubMed/MEDLINE are also covered in the WoS Core Collection.

### *2.2. Methods*

Two dedicated programs, MHNetw.exe[2] and CitNetw.exe,[3] have been developed to generate reference matrices using the PubMed/MEDLine and the WoS data, respectively. The matrices are provided in the Pajek format. CitNetw.exe generates the cited/citing matrix with the citing documents as units of analysis in the rows and the cited references as variables in the columns; MHNetw.exe generates a similar matrix, but with the MeSH in the columns. The number of citing documents is determined by the retrieval from PubMed/MEDLINE or Medline in WoS, respectively. Instructions for how to use the databases and routines are provided in Appendix I.

[2] MHNetw.exe is available from http://www.leydesdorff.net/software/mhnetw for download.
[3] CitNetw.exe is available from http://www.leydesdorff.net/software/citnetw for download.

The routine MHNetw.exe presumes that the data from WoS with the citation information is already organized (by CitNetw.exe) in the same folder so that the citation information can be retrieved locally and attributed to the MeSH categories. If this data is not yet present, the user is first prompted with a search string in the file "string.wos" that can be used at the advanced search interface of WoS.[4]

Both MHNetw.exe and CitNetw.exe provide the following files:

1. "Mtrx.net" contains the reference matrix in the Pajek format; the Pajek format allows for virtually unlimited file sizes.
2. The SPSS syntax file "mtrx.sps" reads the reference matrix ("mtrx.txt") into SPSS and saves this file as an SPSS systems file ("mtrx.sav"). MeSH terms are included as variable labels in the case of MHNetw.exe; in the case of CitNetw.exe, the cited references are the variable labels. The user can combine the two matrices using, for example, Excel.

MHNetw.exe additionally provides:

a) Cr_mh.net, which contains the 2-mode matrix of cited references (CR) in the rows and MeSH terms in the columns;
b) Jcr_mh.net, which simplifies cr_mh.net by using only the abbreviated journal names in the cited references in the rows and MeSH terms in the columns;
c) The file jcr_mh_a.net, which contains the same information (abbreviated journal names and MeSH categories), but organized differently: both CR and MeSH are attributed as variables to the documents under study as the cases (in the rows). Within Pajek, one can convert this matrix into an affiliations matrix (using *Network > 2-Mode Network > 2-Mode to 1-Mode > Columns*). One can also export this file (e.g., to SPSS) for cosine-normalization of the matrix.

[4] One can use this string also for computing the Relative Citation Ratios at https://icite.od.nih.gov/analysis (Hutchins *et al*., 2015). However, this facility has currently a limitation of 200 PubMed identifiers.

CitNetw.exe, furthermore, provides a file "lcs.net" containing the cited/citing matrix for the bounded citation network of the citing documents under study. The bounded citation network corresponds with what was defined as the "local citation environment" in HistCite™ (Garfield, Pudovkin, & Istomin, 2003; Garfield, Sher, & Torpie, 1964). The cited references are matched against a string composed from the meta-data of the citing document using the standard WoS-format of the cited references: "Name Initial, publication year, abbreviated journal title, volume number, and page number" (e.g., "Zhang CL, 2002, CLIN CANCER RES, V8, P1234"). The matrix may be somewhat different from the one obtained from using HistCite™ because of different matching and disambiguation rules.

In order to proceed with main-path analysis in Pajek, the network has to be a-cyclical (de Nooy *et al*., 2011, pp. 244f.). If needed, one can make the network a-cyclical within Pajek by using the following steps in the order specified in Table 1.

**Table 1.** Main or critical path analysis using lcs.net

1. Extract the largest component from the network:
    a. Network > Create partition > Component > Weak
    b. Operations > Network + Partition > Extract subnetwork > Choose cluster;
2. Remove strong components from the largest component:
    a. Network > Create partition > Component > Strong
    b. Operations > Network + Partition > Shrink network > [use default values]
3. Remove loops
    a. Network > Create new network > Transform > Remove > Loops
4. Create main path (or critical path):
    a. Network > Acyclic network > Create weighted > Traversal > SPC
    b. Network > Acyclic network > Create (Sub)Network > Main Paths

The choice of "Main Path > Global Search > Standard", for example, leads to the extraction of the subnetwork with the main path; this subnetwork is selected as the active network. The main path can then be drawn and/or further analyzed.

Note that the cited references are not disambiguated by these routines, but are used as they appear on the input file. The user may wish to disambiguate the references before entering this routine; for example, by using CRExplorer.EXE at http://www.crexplorer.net (Thor, Marx, Leydesdorff, & Bornmann, 2016).

## 3. Results

### *3.1. Descriptive*

Figure 1 shows the number of documents in the set over time and the development of the ratio of citations per publication (*c*/*p*). As noted, the research program under study was triggered by a paper in 1992 (Hardy & Higgins, 1992). However, there are 11 papers in the set with publication dates in 1991 predating this formulation. In the first decade, the number of publications shows exponential growth; but over the full time span linear growth prevails. In other words, this line of research is no longer booming, but since around 2000 can be considered as "normal science."

The *c*/*p* ratio declines linearly with the subsequently shorter citation windows for more recent papers. However, the decline in this ratio may also indicate a diminishing attractiveness of this line of research (Hardy & Selkoe, 2002). The sharp decline in the number of publications in the most recent years confirms this inference (Selkoe & Hardy, 2016). Recently, Herrup (2015) concluded "that the time has come to face our fears and reject the amyloid cascade hypothesis," albeit at the moment without an alternative explanation of Alzheimer's Disease.

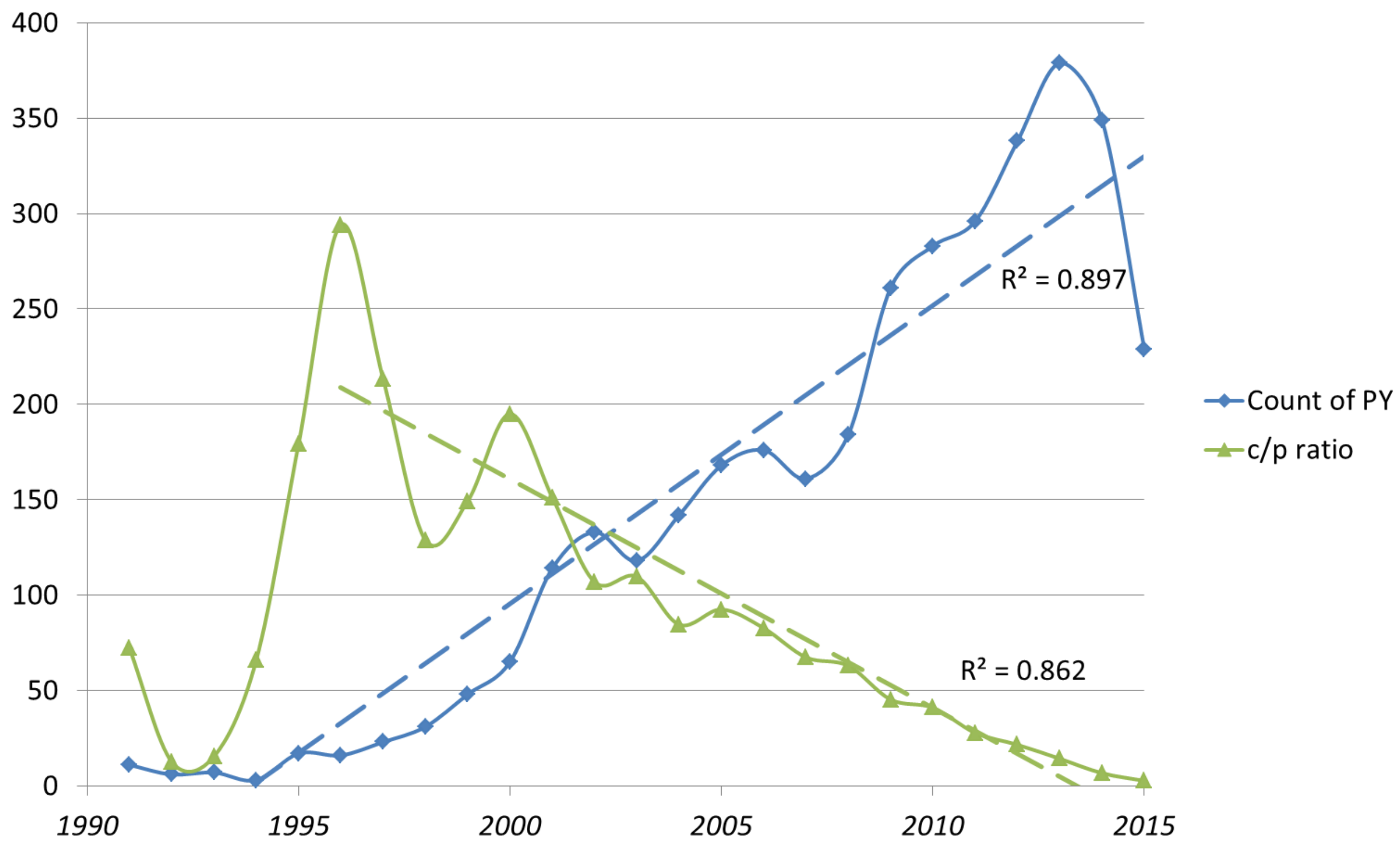

**Figure 1**: Number of yearly papers (♦) and citations per paper ratio (▲) over time.

Table 2 tells us that the number of cited references in the papers under study (176,670) is almost three times that of the MeSH terms attributed (62,648). In terms of unique cited references (67,831) versus unique MeSH terms (3,532), the ratio is further worsened. On a map, the citations would completely overshadow the MeSH terms. However, the number of referenced journals (5,345) is of the same order as the unique MeSH terms.

**Table 2:** Some descriptive statistics of the data under study.

| | **PubMed/MEDLINE** | **WoS** |
|---|---|---|
| *N of documents* | 3,558 | 3,416 |
| *MeSH references* | 62,648 | |
| *Unique MeSH terms* | 3,532 | |
| *Cited references* | | 176,670 |
| *Unique cited references* | | 67,831 |
| *Referenced journals* | | 5,345 |

Figure 2 provides a map which can be generated using the 2-mode matrix of 5,345 abbreviated journal names in the references (red) versus 3,482 MeSH terms (green).[5] (To generate this figure, the file jcr_mh.net was input into Pajek and from there into VOSviewer for the visualization). The figure shows the very central position of the *Journal of Neuroscience* among the references. Although there are more unique references to journals than to MeSH, their concentration indicates that the red-colored journals form a backbone structure with the MeSH terms spreading out as variations. This is the dominant structure in this data: the journals provide a core structure and the MeSH terms the variation. The journals are more concentrated than the MeSH terms (Table 3): the Gini coefficient of the journal distribution is 0.937 while it is 0.852 for the distribution of MeSH.

[5] Fifty of the 3,532 MeSH terms were not related in this case.

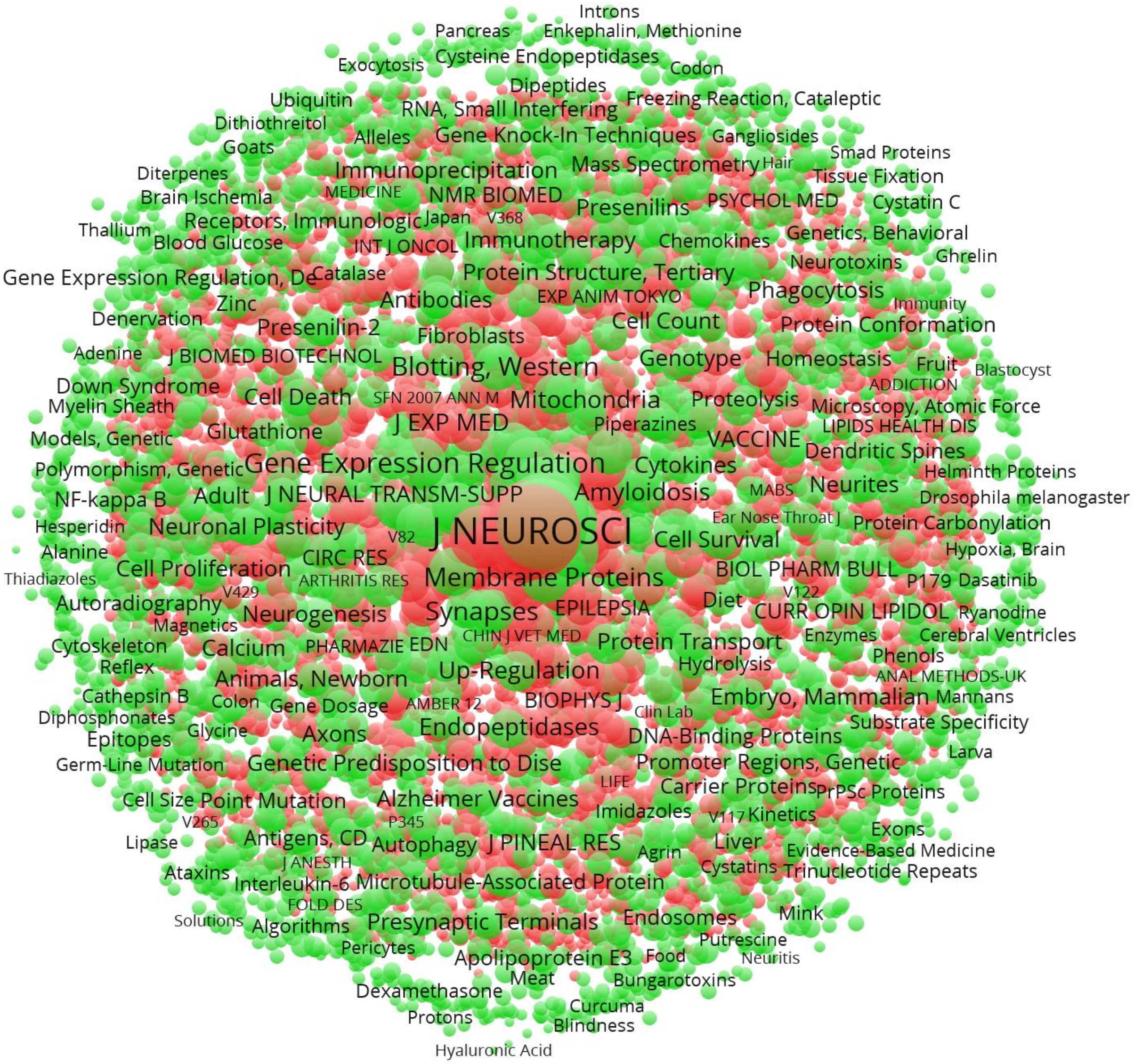

**Figure 2**: Visualization of the 2-mode matrix jcr_mh.net showing 5,345 journals cited and 3,482 MH in 3,558 documents. Layout with Kamada & Kawai (1989); visualization in VOSviewer. This map can be web-started at http://www.vosviewer.com/vosviewer.php?map=http://www.leydesdorff.net/software/mhnetw/jcr_mh_map.txt&label_size_variation=0.3&zoom_level=1&scale=0.9

**Table 3**: Ten most frequently cited journals and ten most frequently referenced MeSH.

| | Referenced Journal | N | MESH | N |
|---|---|---|---|---|
| 1 | *J Neurosci* | 11,842 | Alzheimer Disease | 3,558 |
| 2 | *P Natl Acad Sci USA* | 9,250 | Animals | 3,558 |
| 3 | *J Biol Chem* | 8,616 | Mice | 3,392 |
| 4 | *Nature* | 6,874 | Mice, Transgenic[6] | 3,333 |
| 5 | *Science* | 6,385 | Amyloid beta-Peptides | 2,492 |
| 6 | *Neuron* | 5,428 | Humans | 2,336 |
| 7 | *Neurobiol Aging* | 5,360 | Disease Models, Animal | 2,141 |
| 8 | *J Neurochem* | 4,461 | Amyloid beta-Protein Precursor[5] | 2,090 |
| 9 | *Nat Med* | 3,224 | Brain | 1,374 |
| 10 | *Am J Pathol* | 3,079 | Male | 1,053 |

### *3.2. Analysis and decomposition*

Whereas multivariate analysis (e.g., factor analysis) is limited by systems and software limitations, the new decomposition algorithms enable us to decompose large and even very large matrices. The above matrix (Figure 2), for example, can robustly be decomposed into five clusters using the algorithm of Blondel *et al.* (2008); the modularity of the network is low ($Q$ = 0.066).[7] Figure 3, for example, shows the fourth component consisting of 598 cited journals versus 326 MeSH terms focusing on techniques such as neuro-imaging. This cluster can be further subdivided into nine components ($Q$ = 0.375).

[6] While "Mice, transgenic" and "Amyloid beta-Protein Precursor" were both part of the original search string, the search also retrieves records with MeSH subsumed under these categories: these are "Mice, knockout" (333 times) and "Amyloid beta-Peptides" (2,492 times), respectively.

[7] The decomposition algorithm of VOSviewer distinguishes more than one hundred clusters after symmetrizing the asymmetrical matrix internally by summing the cells (*i*,*j*) and (*j*,*i*).

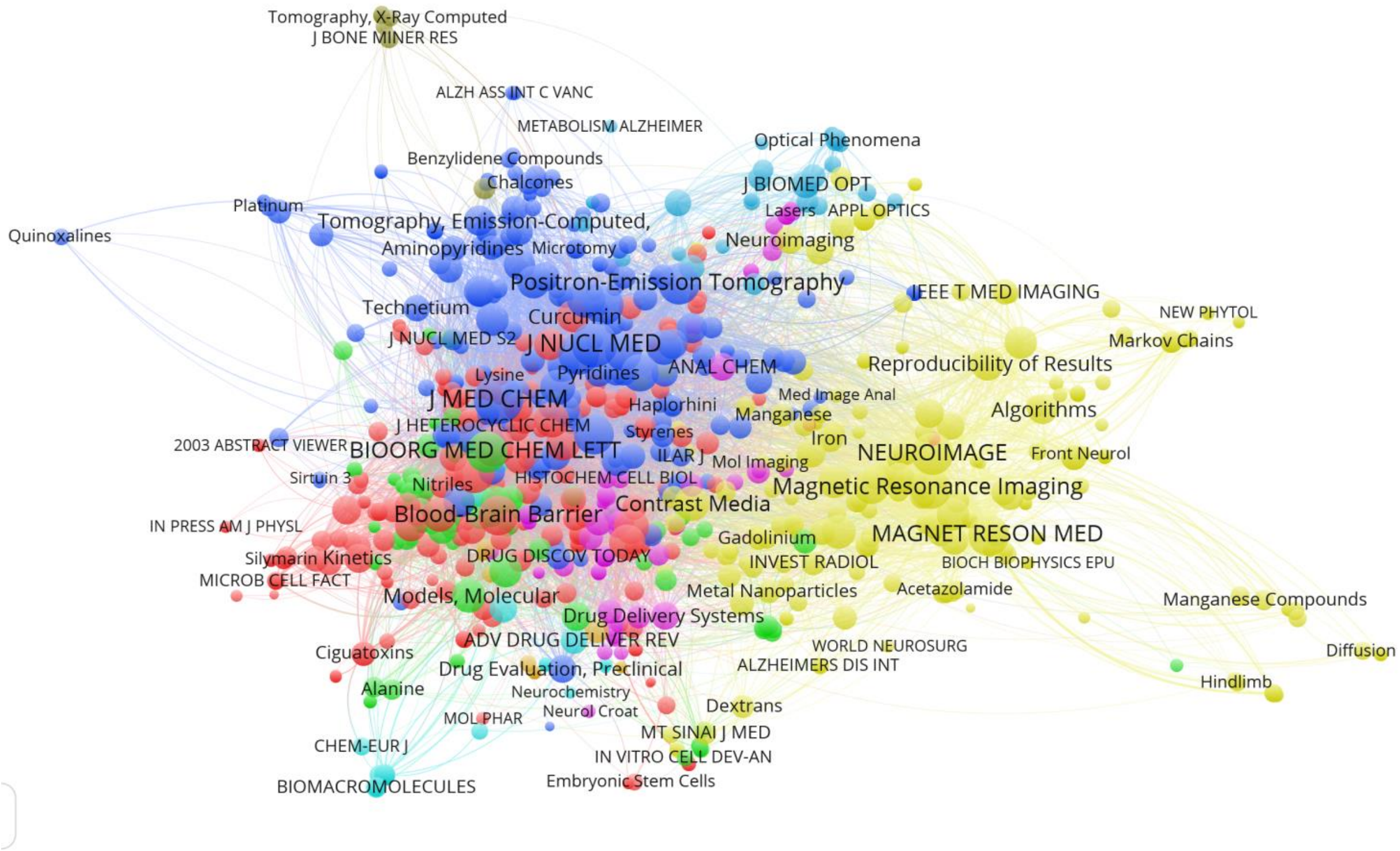

**Figure 3**: Visualization of the fourth component of the 2-mode matrix jcr_mh.net showing 598 journals cited and 326 MH. Nine clusters are distinguished with modularity *Q* = 0.375 (Blondel *et al*., 2008). Layout using (Fruchterman & Reingold, 1991) and visualization in VOSviewer. This map can be web-started at http://www.vosviewer.com/vosviewer.php?map=http://www.leydesdorff.net/software/mhnetw/comp4map.txt&network=http://www.leydesdorff.net/software/mhnetw/comp4net.txt&label_size_variation=0.2&zoom_level=1&scale=1.20&colored_lines&n_lines=10000&curved_lines

The file jcr_mh_a.net organizes the same information as a matrix of the 3,558 documents under study as the cases and both the MeSH terms and abbreviated journal titles as variables in the columns. Using this file, one can normalize the variables or proceed to multivariate analysis. After normalization using the Jaccard index—available in UCInet—the highly centralized structure, indeed, has disappeared. The resulting 1-mode similarity matrix can be decomposed into approximately 70 components by the algorithm of Blondel *et al.* (2008) and into 61 by the algorithm of VOSviewer (Waltman & van Eck, 2012). The modularity is an order of magnitude larger than in the previous case ($Q$ = 0.577). After this normalization, however, journal names come even more to the fore on the map (Figure 3),[8] indicating their structural role in this information.

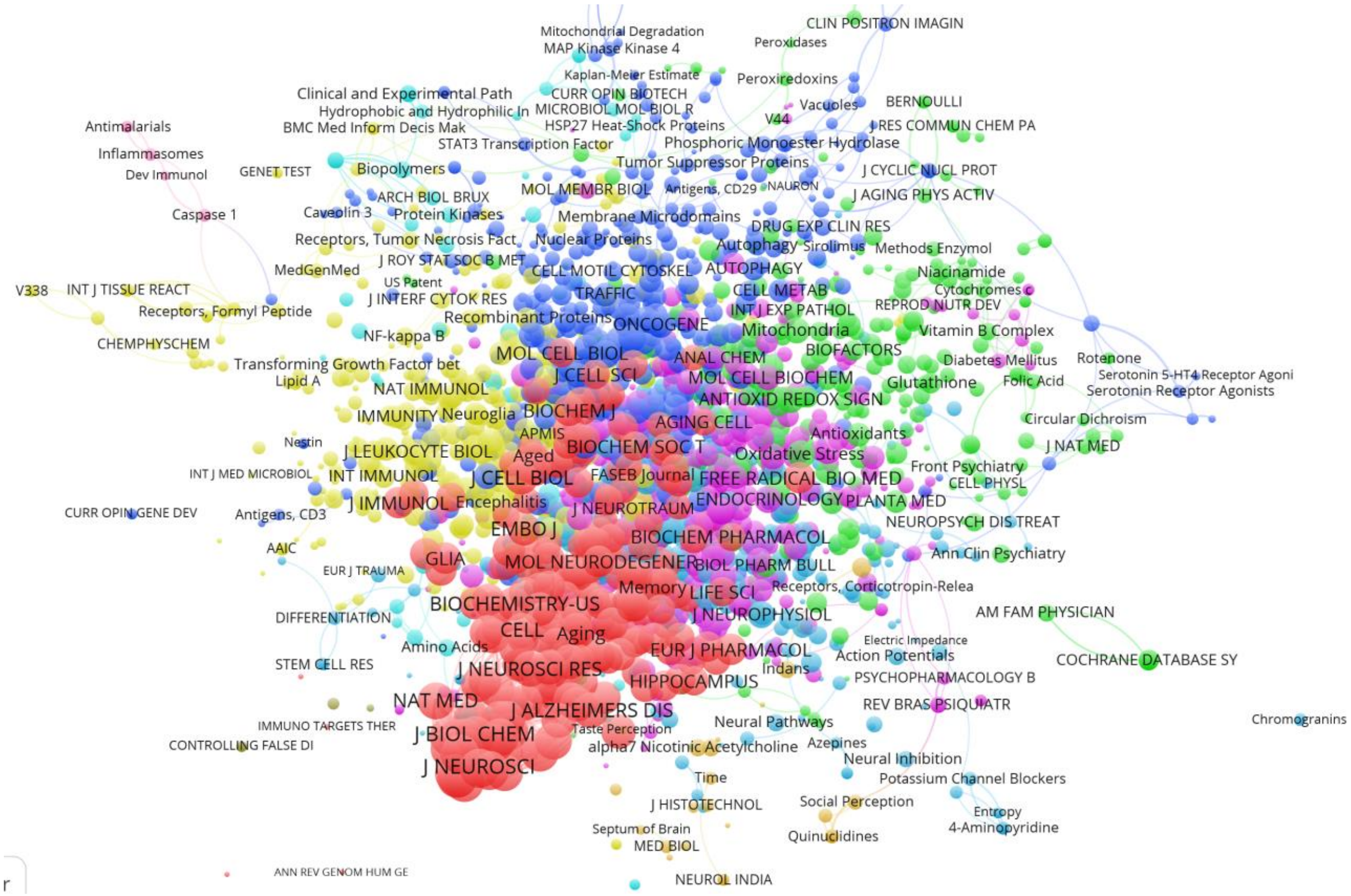

**Figure 4**: First component of the Jaccard-normalized matrix: 1083 cited journals and 900 MeSH terms; subdivided into 11 clusters (Blondel *et al*., 2008; $Q$ = 0.220); layout and visualization using VOSviewer.

In summary, the abbreviated journal names in the references provide us with far greater access to the structure in the matrix than do the MeSH terms. Referenced journals reflect the archival knowledge base on which the new knowledge claims build, whereas MeSH terms position papers as variation (including novelty; Boudreau *et al*., in press) at the research front. The MeSH terms are attributed from the perspective of hindsight. In other words, the MeSH classification which operates at the paper level may be less suited for the normalization of citations than journals or journal categories, which can reveal archival structures.

[8] The full map can be web-started at http://www.vosviewer.com/vosviewer.php?map=http://www.leydesdorff.net/software/mhnetw/jaccard_map.txt&label_size_variation=0.2&zoom_level=1&scale=0.85

### *3.3. Main path analysis*

As noted, CitNetw.exe also generates a file "lcs.net" containing the bounded network of the papers under study with "local citation scores (Garfield *et al*., 2003). Using the instruction provided in section 2, one can generate a main path using Pajek. Figure 5, for example, shows the so-called "key-route main path" as the most recommended option for this analysis (Liu & Lu, 2012). Forty of the 3,416 documents downloaded from WoS (or slightly more than 1%) are located on this main path.[9]

It is beyond the scope of this paper to compare these results with other options for main-path or critical path analysis (Batagelj, 2003; Hummon & Doreian, 1989). A review of the various options is provided by Liu & Lu (2012), who suggest that a combination of the results of several algorithms into an integrated model can improve the quality of the main-path analysis (cf. Lucio-Arias & Leydesdorff, 2008). The resulting main path can be further analyzed as a Pajek file; for example, the colors in Figure 5 show the results of decomposition using the algorithm of Blondel *et al.* (2008).

---

[9] Thirty-four of these documents are located on both the standard and critical main paths as a reduction to a single main path.

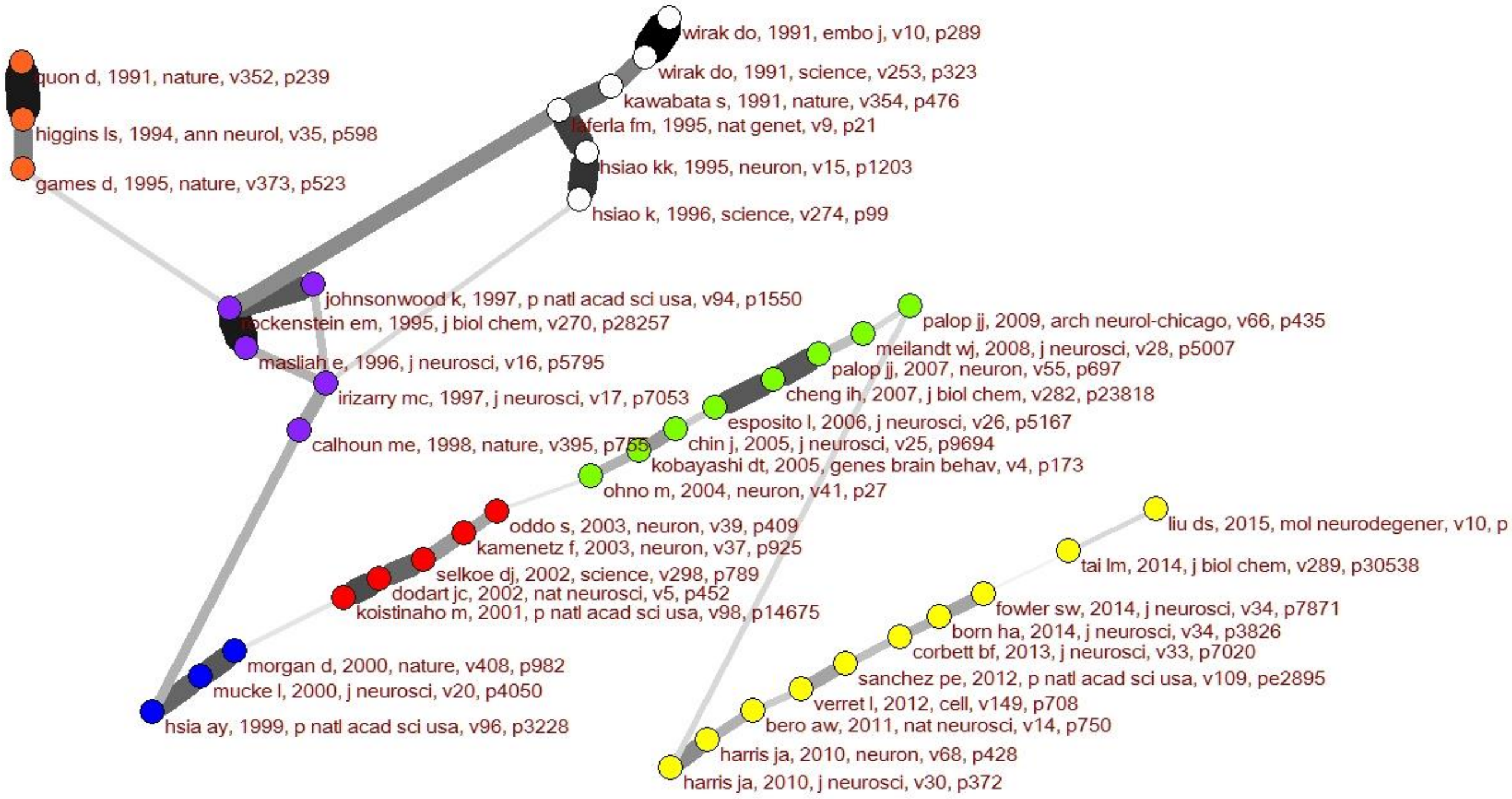

**Figure 5**: Forty papers on the so-called "key route global main path" in the citations among the 3,416 WoS documents under study. Decomposition using the Louvain algorithm in Pajek (Blondel *et al*., 2008; $Q$ = 0.757); layout using Kamada & Kawai (1989).

The generation of a main path of forty articles for a line of investigation encompassing approximately 3,500 papers is appealing due to the reduction by two orders of magnitude in the amount one would need to read to obtain an understanding of this subfield. However, a main path remains an algorithmic construct that one can use heuristically, but that otherwise requires validation. For example, the paper by Kawabata *et al*. (1991) published in December 1991 in *Nature* was retracted on March 19, 1992. This paper received 16 citations by other papers on the main path, thirteen of them in the years after the retraction. From an intellectual perspective, one might consider removing this article from the pool of candidate nodes before regenerating the main path.

The two main scientific awards within the field of AD research are the “Potamkin Prize for Research in Pick’s, Alzheimer’s, and Related Diseases” and the “MetLife Foundation Award for Medical Research in Alzheimer’s Disease.” Both prizes have been awarded since the late 1980s, thus capturing in full the time period of our analysis. Forty investigators have won both awards. The main path (as depicted in Figure 5) includes one or more papers from twelve of these authors.

## Conclusions

We have developed two routines that enable the researcher to generate matrices of citing versus cited documents and/or citing documents versus MeSH terms. The data from WoS and PubMed/Medline was integrated using the PubMed Identifier (PMID). Since the number of citing documents is (almost) the same in both cases, the two matrices can also be juxtaposed and then merged so that combinations of citations and MeSH terms can be analyzed. These combinations can perhaps be considered as hybrid indicators (e.g., Braam, Moed, & van Raan, 1991).

Aggregation of the cited references at the journal level reduces the number of variables by orders of magnitude; the resulting numbers are comparable to the numbers of MeSH categories attributed. Further analysis leads to the conclusion that the abbreviated journal names in the cited references indicate a core structure of the set,[1] whereas the MeSH are attributed regarding to their relevance to current research options. This classification therefore seems less suited for carrying the normalization of citations than journals or journal groups.

[1] The results of a core/periphery analysis are not shown here, but can be web-started at http://www.vosviewer.com/vosviewer.php?map=http://www.leydesdorff.net/software/mhnetw/kcore_map.txt&network=http://www.leydesdorff.net/software/mhnetw/kcore_net.txt&label_size_variation=0.3&zoom_level=2&scale=0.

In the context of this study, main-path analysis provides another example of the research potential of organizing the data into primary matrices extracted from downloads of PubMed and WoS. As a perspective for further research, Hellsten & Leydesdorff (2016), for example, analyze translational research in medicine in terms of combinations of MeSH terms, institutional addresses, and journal names. By considering these and other (meta-)data as attributes of documents, one can merge matrices and combine dimensions in the data as we have done above for cited references and MeSH terms, but also beyond two dimensions in terms of *n*-mode arrays and therefore heterogeneous networks (Callon & Latour, 1981; Law, 1986).

**References:**

Agarwal, P., & Searls, D. B. (2009). Can literature analysis identify innovation drivers in drug discovery? *Nature Reviews Drug Discovery, 8*(11), 865-878.

Batagelj, V. (2003). Efficient algorithms for citation network analysis. *arXiv preprint cs/0309023*.

Boudreau, K. J., Guinan, E., Lakhani, K., & Riedl, C. (2014). Looking across and looking beyond the knowledge frontier: Intellectual distance and resource allocation in science. *Management Science, 0(0), null*. doi: 10.1287/mnsc.2015.2285

Braam, R. R., Moed, H. F., & van Raan, A. F. J. (1991). Mapping of science by combined co-citation and word analysis. I. Structural aspects. *Journal of the American Society for Information Science, 42*(4), 233-251.

Chen, C. (2016, in press). Grand Challenges in Measuring and Characterizing Scholarly Impact. *Frontiers in Research Metrics and Analytics,* (2016; in press)

Fruchterman, T., & Reingold, E. (1991). Graph drawing by force-directed replacement. *Software—Practice and Experience, 21*, 1129-1166.

Garfield, E. (1979). Is citation analysis a legitimate evaluation tool? *Scientometrics, 1*(4), 359-375.

9 . The *k*-core analysis is based on relations with a value of ten or more and confirms that journal names are prevailing in the core set.

Garfield, E., Pudovkin, A. I., & Istomin, V. S. (2003). Why do we need algorithmic historiography? *Journal of the American Society for Information Science and Technology, 54*(5), 400-412.

Garfield, E., Sher, I. H., & Torpie, R. J. (1964). The use of citation data in writing the history of science. Philadelphia, PA: Institute for Scientific Information.

Hardy, J., & Allsop, D. (1991). Amyloid deposition as the central event in the aetiology of Alzheimer's disease. *Trends in pharmacological sciences, 12*, 383-388.

Hardy, J., & Selkoe, D. J. (2002). The amyloid hypothesis of Alzheimer's disease: progress and problems on the road to therapeutics. *Science, 297*(5580), 353-356.

Hellsten, I., & Leydesdorff, L. (2016). *Translational Research in Medicine: Multi-mode network analysis*. Paper presented at the Conference Networks in the Global World, St Petersburg, July 1-3, 2016.

Herrup, K. (2015). The case for rejecting the amyloid cascade hypothesis. *Nature neuroscience*, 794-799.

Hicks, D., & Wang, J. (2011). Coverage and overlap of the new social science and humanities journal lists. *Journal of the American Society for Information Science and Technology, 62*(2), 284-294.

Hummon, N. P., & Doreian, P. (1989). Connectivity in a citation network: The development of DNA theory. *Social Networks, 11*(1), 39-63.

Hutchins, B. I., Yuan, X., Anderson, J. M., & Santangelo, G. M. (2016). Relative Citation Ratio (RCR): A new metric that uses citation rates to measure influence at the article level. *bioRxiv*, 029629.

Leydesdorff, L., Rotolo, D., & Rafols, I. (2012). Bibliometric Perspectives on Medical Innovation using the Medical Subject Headings (MeSH) of PubMed. *Journal of the American Society for Information Science and Technology, 63*(11), 2239-2253. doi: 10.1002/asi.22715

Liu, J. S., & Lu, L. Y. (2012). An integrated approach for main path analysis: Development of the Hirsch index as an example. *Journal of the American Society for Information Science and Technology, 63*(3), 528-542.

Lucio-Arias, D., & Leydesdorff, L. (2008). Main-path analysis and path-dependent transitions in HistCite™-based historiograms. *Journal of the American Society for Information Science and Technology, 59*(12), 1948-1962.

Lundberg, J., Fransson, A., Brommels, M., Skår, J., & Lundkvist, I. (2006). Is it better or just the same? Article identification strategies impact bibliometric assessments. *Scientometrics, 66*(1), 183-197.

Petersen, A., Rotolo, D., & Leydesdorff, L. (2016). A Triple Helix Model of Medical Innovations: *Supply, Demand,* and *Technological Capabilities* in Terms of Medical Subject Headings. *Research Policy, 45*(3), 666-681. doi: 10.1016/j.respol.2015.12.004

Reitz, C. (2012). Alzheimer's Disease and the Amyloid Cascade Hypothesis: A Critical Review. *International Journal of Alzheimer's Disease, 2012*(Article ID 369808), 11 pages; http://dx.doi.org/10.1155/2012/369808.

Rotolo, D., & Leydesdorff, L. (2015). Matching MEDLINE/PubMed Data with Web of Science (WoS): A Routine in R language. *Journal of the Association for Information Science and Technology 66*(10), 2155-2159. doi: 10.1002/asi.23385

Rotolo, D., Rafols, I., Hopkins, M. M., & Leydesdorff, L. (2016). Strategic intelligence on emerging technologies: Scientometric overlay mapping. *Journal of the Association for Information Science and Technology*, Early view.

Selkoe, D. J. (1991). The molecular pathology of Alzheimer's disease. *Neuron, 6*(4), 487-498.

Selkoe, D. J., & Hardy, J. (2016). The amyloid hypothesis of Alzheimer's disease at 25 years. *EMBO molecular medicine, 8*(6), 595-608.

Thor, A., Marx, W., Leydesdorff, L., & Bornmann, L. (2016). Introducing CitedReferencesExplorer : A program for Reference Publication Year Spectroscopy with Cited References Disambiguation. *Journal of Informetrics, 10*(2), 503-515. doi: 10.1016/j.joi.2016.02.005

Waltman, L., & van Eck, N. J. (2012). A new methodology for constructing a publication-level classification system of science. *Journal of the American Society for Information Science and Technology, 63*(12), 2378-2392.

## Appendix I

The two routines CitNetw.EXE and MHNetw.EXE (available at http://www.leydesdorff.net/software/citnetw/) can be used for making complete matrices at the article level in the Pajek and SPSS formats for the analysis of citations and medical subject headings, respectively. The two matrices can also be combined.

On the basis of a download of Web-of-Science data, CitNetw.EXE can generate the citation matrix with the citing papers in the rows and cited references in the columns in the following formats: (*i*) mtrx.net in the Pajek format and (*ii*) mtrx.sps + mtrx.txt for SPSS. The matrix is binary, asymmetrical, 2-mode, and directed. (If so wished, one can transpose this matrix in Pajek or SPSS.) One can process the file "mtrx.net" further in Pajek, UCInet, or Gephi, etc. The file lcs.net (output of CitNetw.Exe) contains the bounded network of citations among the documents under study. This file can be used, for example, for main path analysis (see Appendix II).

Input to both routines is a file "data.txt" containing downloads from WoS and Medline, respectively, in the "plain text" or "Medline" format (tagged). This file is first processed into a format for relational database management. (One is prompted for skipping this reorganization if it was already done in a previous round.) If one wishes to combine the outputs of the two routines, the files mtrx.* should first be saved and stored elsewhere, since these files are overwritten in subsequent runs.

The objective of using MHNetw.EXE is to combine Medical Subject Headings (MeSH) and citation information at the article level. The MeSH are first retrieved from the PubMed database and can be organized into relational data using the routine pubmed.exe at http://www.leydesdorff.net/pubmed . Note that one also needs the file <pubmed.dbf> to be present in the same folder as the data and pubmed.exe. Alternatively, one can retrieve the data from Medline in WoS. The advantage of retrieval from PubMed above retrieval from WoS is that there is no limitation of 500 records each time. The data from either source has first to be organized in the same folder using PubMed.Exe. The program prompts with a question about either source. Input data have always to be named "data.txt".

Output of MHNetw.exe is:

- mtrx.net (Pajek) and mtrx.sps (for SPSS) containing the citing papers as rows and the MeSH as variables in the columns (analogous to CitNetw.exe).
- A file called "string.wos" which contains the search string for obtaining citation information at Web of Science (advanced search).
- The citation scores are written into the file with article descriptors ti.dbf in a field "tc"; citation scores are summed for MeSH into mh1.dbf.
- The file "string.wos" can be used to generate the corresponding file in the Science Citation indices of WoS; the file "string.pubmed" contains analogously the search string if one has worked from the WoS interface.

- The file cr_mh.net contains the citation information (cited references, CR) in the rows and the medical subject headings (MH) in the columns. The cell values provide the number of documents in which cited references and MeSH co-occur.
- The file jcr_mh.net contains the abbreviated journal names in the cited references (CR) in the rows and the medical subject headings (MH) in the columns. The cell values provide the number of documents in which the cited journals and MeSH co-occur.
- The file jcr_mh_a.net contains the same information (abbreviated journal names and MeSH categories), but differently organized: both are attributed as variables to the documents under study as the cases. Within Pajek, one can convert this matrix into an affiliations matrix (using *Network > 2-Mode Network > 2-Mode to 1-Mode > Columns*). One can also export this file to SPSS for cosine-normalization of the matrix.

The asterisks in MeSH terms are discarded in this version. All files operate only on files present in the same folder. Note that mtrx.net, mtrx.txt, and mtrx.sps are overwritten in each run of MHNetw.exe or CitNetw.exe. One is advised to save all files mtrx.* elsewhere or to rename them for this reason.

We suggest the following order of the routines:

1. Download data at PubMed from the user interface at http://www.ncbi.nlm.nih.gov/pubmed/advanced . At the results page thereafter, select under “Send to” the format option MEDLINE and download to a file which has to be (re)named “data.txt”;
2. Run pubmed.exe (with this file data.txt as input) in the presence of pubmed.dbf; both files are available at http://www.leydesdorff.net/pubmed/index.htm ;
3. Use the resulting string “string.wos” at the advanced user interface of WoS; save the retrieval via “Marked list” in portions of 500 records. Combine the data into a file data.txt.
4. Run CitNetw.EXE; save the citation matrices in the files mtrx.* elsewhere;
5. Run MHNetw.EXE; save the matrices that one wishes to use for further analysis. This analysis may take long.

Appendix 2
**Main Path analysis**

Alongside other files, CitNetw.EXE generates a file lcs.net containing the citations within the bounded domain of the document set(s) under study. (This domain corresponds to the so-called local citation scores (lcs) in HistCite™.) However, the cited references are not disambiguated, but used as they are provided by WoS. The user may wish to disambiguate the references before entering this routine (for example, by using CRExplorer.EXE.) The cited references are matched against a string composed from the citing document using the WoS-format of the cited references "Name Initial, publication year, abbreviated journal title, volume number, and page number" as follows: "Zhang CL, 2002, CLIN CANCER RES, V8, P1234".

The output file **lcs.net** contains a matrix with the citing documents in the rows and the cited ones in the columns. The matrix may be somewhat different from the one which one can obtain from using HistCite™ because of different matching and disambiguation procedures.

In order to proceed with main-path analysis in Pajek, the network has to be made a-cyclical (de Nooy *et al*., 2011, pp. 244f.). One can make the network a-cyclical within Pajek using the following steps *in this order*:

1. Extract the largest component from the network:
    a. Network > Create partition > Component > Weak
    b. Operations > Network + Partition > Extract subnetwork > Choose cluster 1;
2. Remove strong components from the largest component:
    a. Network > Create partition > Component > Strong
    b. Operations > Network + Partition > Shrink network > [use default values]
3. Remove loops:
    a. Network > Create new network > Transform > Remove > Loops
4. Create main path (or critical path):
    a. Network > Acyclic network > Create weighted > Traversal > SPC
    b. Network > Acyclic network > Create (Sub)Network > Main Paths

The subsequent choice among the options of Main Path for "> Global Search > Standard", for example, leads to the extraction of the subnetwork with the main path; this subnetwork is selected as the active network. The main path can then be drawn and/or further analyzed.